\documentclass[prl,final,twocolumn,showpacs]{revtex4}

\usepackage{graphicx}% Include figure files
\usepackage{dcolumn}% Align table columns on decimal point

\begin{document}
\title{Conductance properties of nanotubes coupled to superconducting leads:
signatures of Andreev states dynamics}
\author{E. Vecino$^1$, M.R. Buitelaar$^{2*}$, A. Mart\'{\i}n-Rodero$^1$, C.
Sch\"onenberger$^2$ and A. Levy Yeyati$^1$}
\affiliation{$^1$Departamento de F\'\i sica Te\'orica de la Materia
Condensada C-V. \\
Universidad Aut\'onoma de Madrid. E-28049 Madrid. Spain.}
\affiliation{$^2$Institut f\"ur
Physik, Universit\"at Basel, Klingelbergstrasse
82, CH-4056 Basel, Switzerland.}
\date{\today}

\begin{abstract}
We present a combined experimental and theoretical analysis of the
low bias conductance properties of carbon nanotubes coupled to superconducting
leads. In the Kondo regime the conductance exhibits a zero bias peak
which can be several times larger than the unitary limit in the normal
case. This zero bias peak can be understood by analyzing the dynamics of
the subgap Andreev states under an applied bias voltage. It is shown
that the existence of a linear regime is linked to the presence of a
finite relaxation rate within the system. The theory provides a good
fitting of the experimental results.
\end{abstract}


\maketitle

Carbon nanotubes connected to metallic electrodes
allow to study many different regimes in mesoscopic electron transport
exhibiting either ballistic or diffusive behavior \cite{ballistic,diffusive};
Luttinger liquid features \cite{luttinger}, etc.
Several experiments \cite{qdots} have demonstrated that carbon nanotubes weakly
coupled to normal leads can also exhibit Coulomb blockade and
Kondo effect, i.e. the characteristic physics of quantum dots (QD).
Very recently the research on the transport properties of this system
has been extended with the inclusion of superconductivity on the leads
\cite{super-leads}.
In particular Buitelaar et al. \cite{basel1} have achieved a
physical realization of a S-QD-S system using
carbon nanotubes coupled to Al/Au leads.
Their results, showing an enhancement of the
conductance in the Kondo regime with respect to the normal case, have
attracted considerable theoretical attention \cite{basel2,us03,kang03}.

From a theoretical point of view transport in a voltage biased
S-QD-S system poses a challenging problem due to the interplay
between electron correlation effects and multiple Andreev
reflection (MAR) processes which provide the main mechanism for
quasiparticle transport at small bias. This problem has been
addressed by several authors at different levels of approximation.
Thus, in Refs. \cite{us97} numerical results were presented for
the S-QD-S system neglecting electron correlation effects. This
approach has also been used by Buitelaar et al. \cite{basel2} for
analyzing the experimental results for the subgap structure of
Ref. \cite{basel1}. On the other hand, Avishai et al.
\cite{avishai} have introduced the effect of electron interactions
at the level of the slave boson mean field approach and obtained
numerical results for the current and the noise spectrum in the
Kondo regime. More recently, some of us have shown that the
conductance properties of a S-QD-S system in the Kondo regime can
be understood in terms of the dynamics of subgap states (Andreev
states) under an applied bias voltage \cite{us03}.

\begin{figure}
\includegraphics[width =\columnwidth]{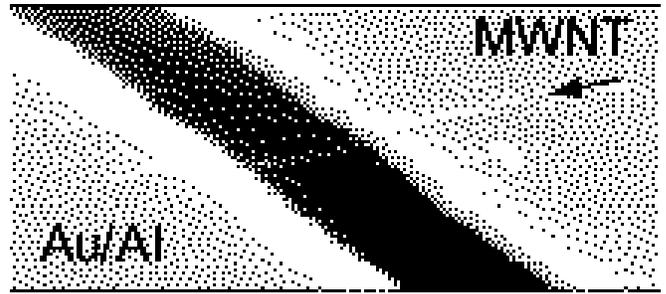}
\caption{Typical 2-terminal device geometry. For the measurements
presented here, the electrode spacing is 0.25 $\mu$m and the MWNT
length 1.5 $\mu$m. The (oxidized) Si substrate is used as a gate
electrode.} \label{device}
\end{figure}

In the present work we combine experimental and theoretical
analysis in order to identify signatures of Andreev states
dynamics in the low bias conductance of carbon nanotubes coupled
to superconducting leads. We shall first describe the setup used for the
conductance measurements and discuss the main experimental results.
We then introduce the theoretical model in which we assume that
the system can be appropriately analyzed as a single level quantum dot.
In contrast to Ref. \cite{us03}, in the present calculations we
explicitly take into account the presence of a finite inelastic
relaxation rate within the system. As will be shown the low bias
conductance properties are extremely sensitive to this rate.
The comparison with the experimental results allows to determine the
size of this parameter in the actual system
as well as to confirm the main predictions of the theory.

The device we consider consists of an
individual multiwall carbon nanotube (MWNT) of 1.5 $\mu$m length
between source and drain electrodes that are separated by 250 nm,
Fig.\ref{device}. The lithographically defined leads were
evaporated over the MWNT, 45 nm of Au followed by 135 nm of Al.
The degenerately doped Si substrate was used as a gate electrode.

\begin{figure}
\includegraphics[width = 7cm]{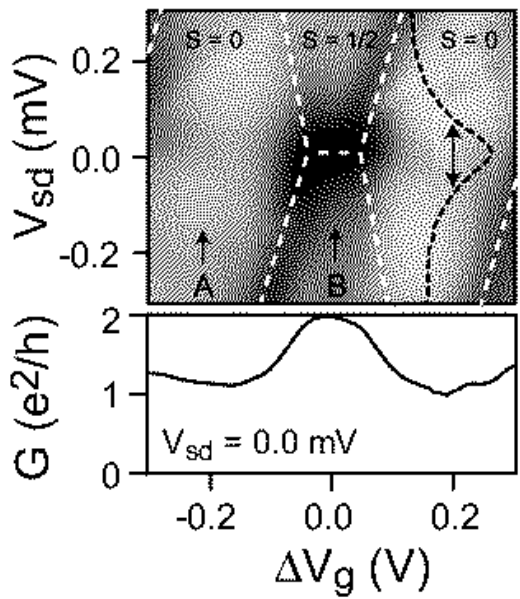}
\caption{Gray scale representation of the differential conductance
$dI/dV$ versus source-drain ($V_{sd}$) and gate voltage ($V_g$)
with the leads in the normal state (darker = more conductive). The
dashed white lines outline the Coulomb blockade diamonds and the
Kondo ridge. The black dashed line trace is the $dI/dV$ in the
middle of the Kondo ridge (position $B$), from which $T_K$ can be
estimated. The line trace in the bottom graph is the
linear-response conductance.} \label{grayscale}
\end{figure}

The device is first characterized with the contacts driven normal
by applying a small magnetic field of $B = 26$ mT. The critical
field of the electrodes was experimentally determined to be $\sim
12$ mT. Figure \ref{grayscale} shows a gray scale representation
of the measured differential conductance $dI/dV_{sd}$ versus
source-drain ($V_{sd}$) and gate voltage ($V_g$) at $T = 50$ mK
with the leads in the normal state. Measured over a large range of
$V_g$ we observe a series of truncated low-conduction Coulomb
blockade diamonds linked by narrow ridges of high conductance of
which Fig.\ref{grayscale} is a small part, see also
Ref.\cite{basel1}. From these and other Coulomb diamonds at
different $V_g$ we obtain an average charging energy $U_C = 0.45$
meV and an average level spacing $\delta \epsilon = 0.45$ meV,
with $\delta \epsilon$ varying between $0.3$ and $0.7$ meV.

The high conductance ridge around $V_{sd} = 0$ mV in
Fig.\ref{grayscale} is a manifestation of the Kondo effect
occurring when the number of electrons on the dot is odd and the
total spin $S = 1/2$. At $T = 50$ mK the Kondo state seems to be
fully developed and has reached the unitary limit with a
conductance maximum of $2 e^2/h$. The Kondo temperature $T_K$ can
be estimated from the width of the Kondo ridge out of equilibrium
\cite{Meir}. The full width at half maximum (FWHM) corresponds to
$\sim k_B T_K$ which yields a Kondo temperature for the ridge in
Fig.\ref{grayscale} of about 1.86 K. In total we have observed 12
Kondo ridges with Kondo temperatures varying between 0.52 and 3.34
K.

\begin{figure}
\includegraphics[width =\columnwidth]{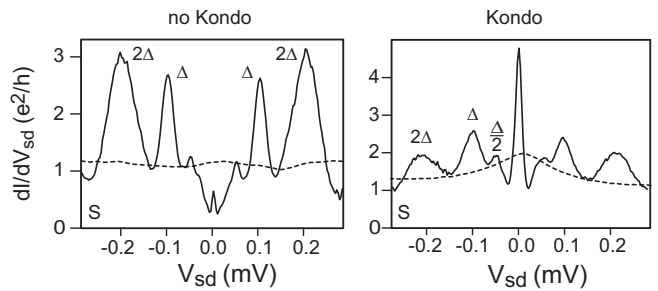}
\caption{Differential conductance $dI/dV$ in the middle of a
Coulomb diamond (left) and in the middle of the Kondo ridge
(right). These correspond to positions $A$ and $B$ in
Fig.\ref{grayscale}, respectively. The difference between the
superconducting state of the leads (solid line) and normal state
(dashed line) is clearly observed. The peaks in the $dI/dV$ in the
superconducting state are due to MAR.} \label{super}
\end{figure}

When the magnetic field is off, the leads are superconducting and
the conductance pattern is dramatically different. This is
illustrated by Fig.\ref{super} which compares $dI/dV$ in the
normal (dashed lines) and superconducting state (solid lines) at
the $V_g$ positions $A$ and $B$ of Fig.\ref{grayscale}. Whereas
the $dI/dV$ in the normal state has no pronounced features (except
for the Kondo resonance at position $B$), a series of peaks appear
in the superconducting state. These are the signature of MAR. The
peaks at $\pm 0.2$ mV correspond to the superconducting gap of $2
\Delta$ and mark the onset of direct quasiparticle tunneling. The
peaks at lower $V_{sd}$ involve one or more Andreev reflections.
The observed subharmonic gap structure is discussed in more detail
in Ref.\cite{basel2}. The superconducting gap energy of $\Delta =
0.1$ meV is slightly less than the expected 0.18 meV for bulk
aluminum which we attribute to the intermediate Au layer
(necessary to obtain good electrical contact to the MWNT).

Whereas the conductance at $V_{sd} = 0$ mV in the superconducting
state is less than its normal state value in the middle of an
$S=0$ Coulomb diamond, the situation is very different for the
Kondo region. Here the conductance increases significantly and a
resonance remains around $V_{sd} = 0$, albeit narrower than in the
normal state. This has been observed for all Kondo resonances,
provided $T_K \gtrsim \Delta$, see Ref.\cite{basel1}. As stated in
the introduction, here we focus on the low-bias region of those
Kondo resonances in the superconducting state and try to identify
signatures of Andreev states dynamics.

To describe theoretically a S-nanotube-S system we shall first
assume that the level spacing $\delta \epsilon$ in the nanotube is
large compared to the superconducting gap in the leads $\Delta$.
In the actual case of the present experiments $\Delta \simeq 0.1$
meV and $\delta \epsilon \simeq $ 0.45 meV, which means that this
approximation is roughly correct. The problem can thus be
mapped into a single-level Anderson model with a model Hamiltonian
$H = H_L + H_R + H_T + H_D$, where $H_{L,R}$ describe the left and
right leads as BCS superconductors, $H_T = \sum_{k, \sigma, \mu=
L,R} t_{\mu} \hat{c}^{\dagger}_{k,\sigma} d_{\sigma} +
\mbox{h.c}$, is the term coupling the dot to the leads,
$\hat{d}^{\dagger}_{\sigma}$ and  $\hat{c}^{\dagger}_{k,\sigma}$
being creation operators for electrons in the dot and in the leads
respectively. $H_D = \sum_{\sigma} \epsilon_0 \hat{n}_{\sigma} + U
\hat{n}_{\uparrow} \hat{n}_{\downarrow}$ is the uncoupled dot
Hamiltonian characterized by the dot level position $\epsilon_0$
and the Coulomb interaction $U$. It is assumed that the coupling
of the dot to the leads in the normal state can be described by
energy independent tunneling rates $\Gamma_{L,R}$.

A crucial point in the analysis of this model is how to deal with the
electron correlation effects. In Ref. \cite{vecino}
a perturbative approach was used to study the zero-voltage case. It was
found that the AS's in the Kondo regime $(T_K > \Delta)$ have essentially
the same phase dependence as in the non-interacting case but with a
reduced amplitude. This is consistent with a Fermi liquid
description of the normal state with renormalized
parameters $\epsilon^*_0$ and $\Gamma_{L,R}^*$,
where $\Gamma_L^* + \Gamma_R^* = 2 \Gamma^* \simeq T_K$ fixes
the width of the Kondo resonance. Although different approximations
differ in the way to relate these quantities to the bare
parameters, the description in terms of the renormalized ones
can be considered as universal \cite{comment:slave-boson}.
In the opposite limit ($T_K < \Delta$), the system is well described
within the Hartree-Fock approximation \cite{varias,vecino,avishai}.

This fully phase-coherent picture is limited in the actual system
by the presence of inelastic scattering mechanisms not considered
in the present model like electron-phonon interactions. We shall later discuss
its effect in the quasiparticle current by introducing a phenomenological
inelastic relaxation rate $\eta$.

In the zero-bias limit the Andreev states are determined by the poles of
the dot retarded Green function which, in the approximation mentioned above
can be written as

\begin{widetext}
\begin{eqnarray}
G^r_{D}(\omega) \simeq \frac{\Gamma^*}{\Gamma}
\left( \begin{array}{cc} \omega -
\epsilon^*_0 - 2 \Gamma^* g^r & 2 \Gamma^* f^r \cos{\phi/2} +
i \delta \Gamma^* \sin{\phi/2} \\
2 \Gamma^* f^r \cos{\phi/2} - i \delta \Gamma^* \sin{\phi/2}
& \omega + \epsilon^*_0 - 2 \Gamma^* g^r
\end{array}
\right)^{-1} ,
\label{dot-retGF}
\end{eqnarray}
\end{widetext}
with $\delta \Gamma^* = \Gamma^*_L - \Gamma^*_R$,
$g^r = -(\omega+i0^+)/\sqrt{\Delta^2 - (\omega+i0^+)^2}$ and
$f^r = \Delta/\sqrt{\Delta^2 - (\omega+i0^+)^2}$ corresponding to the
dimensionless Green functions of the uncoupled electrodes.
In the electron-hole symmetric case ($\epsilon^*_0 = 0$ and
$\Gamma^*_L = \Gamma^*_R$) the AS's are then determined by the equation

\begin{eqnarray}
\omega_s \pm \Delta \cos{\phi/2} +
\frac{\omega_s \sqrt{\Delta^2 - \omega_s^2}}{2 \Gamma^*} = 0 .
\label{determinant}
\end{eqnarray}

\begin{figure}
\includegraphics[width =\columnwidth]{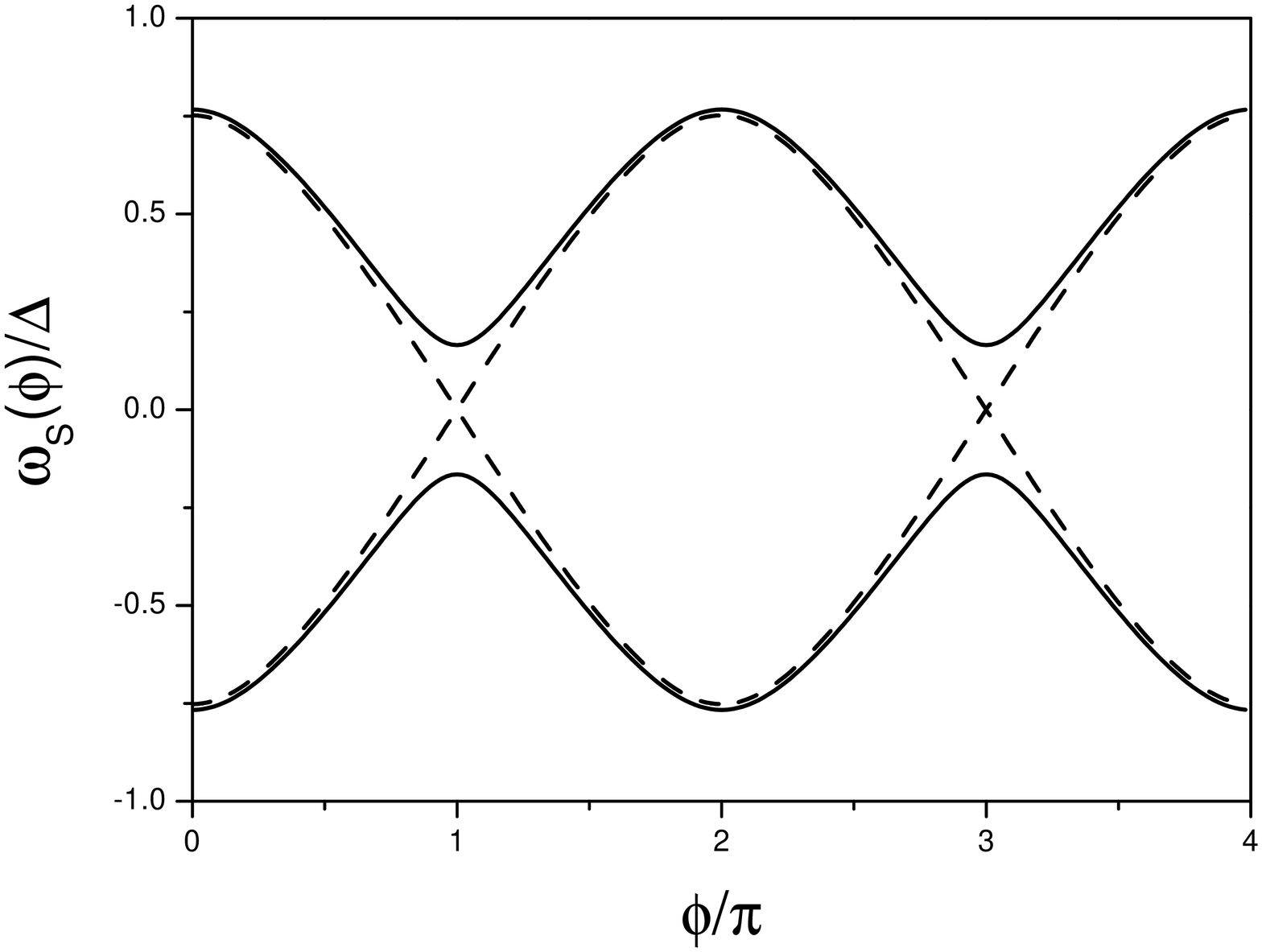}
\caption{Andreev states for a S-QD-S in the Kondo regime.
Dashed and full lines correspond to a symmetric and non-symmetric
situation respectively.}
\label{states}
\end{figure}

The solutions of Eq. (\ref{determinant}) are plotted in Fig.
\ref{states}. As can be observed the AS detach from the continuous
spectrum and behave approximately as
$\omega_s \sim \pm \widetilde{\Delta} \cos{\phi/2}$ with
$\widetilde{\Delta} < \Delta$. The energy interval between the AS and
the edges of the gap is given by $\Delta - \widetilde{\Delta} \simeq
2 \Delta^3/(2 \Gamma^*)^2$ for $\Gamma^* > \Delta$. On the other hand at
the crossing between the two AS's ($\phi=\pm \pi$) $\widetilde{\Delta}$
can be approximated by $\Delta/(1 + \Delta/2\Gamma^*)$.

In a non-symmetric situation ($\epsilon^*_0 \neq 0$ and/or $\Gamma^*_L
\neq \Gamma^*_R$) the
two ballistic states are coupled and
there appear an upper and a lower bound state with an internal gap
of the order of $2 \Delta \sqrt{1-\tau}$, where
$\tau = 4 \Gamma_L^*\Gamma_R^*/\left[(\epsilon^*_0)^2+(\Gamma^*)^2\right]$
is the normal transmission at the Fermi
energy. This is similar to the behavior of the AS's in a point contact
with finite reflection probability, except for the detachment
from the continuous spectrum (see Fig. \ref{states}).

In the voltage biased case it can be assumed that the description in
terms of the renormalized parameters $\epsilon^*_0$ and $\Gamma_{L,R}^*$
remains valid in the regime $eV \ll \Delta \sim T_K$ in which we
are interested.
Within the present model the current operator between the leads and the
dot is given by

\begin{eqnarray}
\hat{I}_{\mu} = \frac{ie}{\hbar} \sum_{k, \sigma}
\left(t_{\mu} \hat{c}^{\dagger}_{k,\sigma} d_{\sigma} - \mbox{h.c} \right)
\end{eqnarray}

For calculating the average current we use the Keldysh
formalism following previous works \cite{tocho}. The main quantities to
determined are the Keldysh Green functions in Nambu space given by
$\hat{G}^{+-}_{ij}(t,t') = i <\hat{\psi}^{\dagger}_j(t') \hat{\psi}_i(t)
>$, where $\hat{\psi}_i$ are the
field operators in Nambu space defined as

\begin{equation}
\hat{\psi_{i}} = \left(
\begin{array}{c}
c_{i \uparrow} \\ c^{\dagger}_{i \downarrow}
\end{array} \right) \hbox{  ,     } \hat{\psi}^{\dagger}_{i}=
\left(
\begin{array}{cc}
 c^{\dagger}_{i \uparrow} & c_{i \downarrow}
\end{array} \right)
\end{equation}

The average current can then be written as

\begin{eqnarray}
I_\mu (t) = \frac{e}{\hbar} \mbox{Tr} \left\{ \hat{\sigma}_z
\left[ \hat{t}_{\mu}
\hat{G}^{+-}_{D,\mu} - \hat{t}^{\dagger}_{\mu} \hat{G}^{+-}_{\mu,D}
\right] \right\}
\end{eqnarray}
where $\hat{t}_{\mu} = t_{\mu} \hat{\sigma}_z$.
One can integrate out the leads to find an expression of the current
only in terms of the  dot Keldysh Green function
$\hat{G}^{+-}_D$, which is
related to the retarded and advanced Green functions by

\begin{eqnarray}
\hat{G}_{D}^{+-} =
\hat{G}_{D}^r \hat{t}_{L} \hat{g}^{+-}_{L}
\hat{t}^{\dagger}_{L} \hat{G}_{D}^a +
\hat{G}_{D}^r \hat{t}_{R} \hat{g}^{+-}_{R}
\hat{t}^{\dagger}_{R} \hat{G}_{D}^a
\label{gpm}
\end{eqnarray}
where $\hat{g}^{+-}_{L,R}$ are the Keldysh Green functions for the
uncoupled electrodes, whose Fourier transform is given by
$n^{L,R}_F(\omega) \left[\hat{g}^a(\omega) - \hat{g}^r(\omega)\right]$,
$n^{L,R}_F(\omega)$
being the corresponding Fermi factor. In Eq. (\ref{gpm})
integration over internal time arguments is implicitly assumed.

Using the different relations between Green functions in Keldsyh space
the current between the dot and the left lead can be written as

\begin{eqnarray}
I_{\mu} (t,V) = \frac{2e}{\hbar} \mbox{Re}
\mbox{Tr} \left\{ \hat{\sigma}_z \left[ \hat{\Gamma}^r_{\mu}
\hat{g}^{+-}_{\mu} + \hat{\Gamma}^{+-}_{\mu} \hat{g}^a_{\mu} \right] \right\}
\label{current}
\end{eqnarray}
where $\hat{\Gamma}_{\mu} = \hat{t}_{\mu} \hat{G}_D \hat{t}^{\dagger}_{\mu}$.
In order to evaluate the current under fixed bias voltage
a double Fourier transformation of the temporal
arguments in the dot Green functions is performed, which leads to a
set of algebraic equations for the different Fourier components
\cite{tocho,us97}. This procedure allows to evaluate the
current as $I(t,V) = (I_L - I_R)/2 =
\sum_n I_n(V) exp{[in\phi(t)]}$, where $\phi(t) =
2eVt/\hbar + \phi_0$. In the present case we shall be interested
in the dc component $I_0(V)$.

\begin{figure}
\includegraphics[width =10cm]{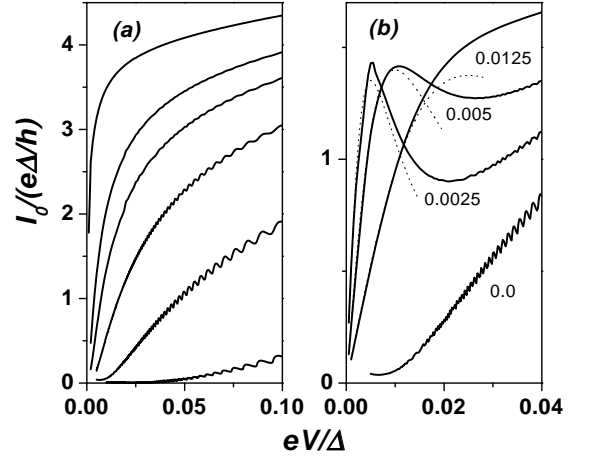}
\caption{Low bias current-voltage characteristic for the symmetric case.
Panel (a) corresponds to the fully phase-coherent case ($\eta=0$) and
different values of $\Gamma^*/\Delta = 10$, 5, 4, 3, 2, 1 from top to
bottom. Panel (b) corresponds to $\Gamma^*/\Delta = 2$ and different
values of $\eta/\Delta$ as indicated. The dotted lines show the analytic
description of the $eV < \eta$ regime given by Eq. (\ref{analytical-dc}).}
\label{ivteorica}
\end{figure}

The behavior of the dc current at low bias is shown in Fig.
\ref{ivteorica}. When $\Gamma^* \gg \Delta$
the $IV$ characteristics of a ballistic quantum point contact is
gradually recovered.
As $\Gamma^*/\Delta$ is reduced the low bias conductance is
gradually suppressed.
When $\Gamma^* \sim \Delta$, MAR oscillations with a period $\Delta/n$, where
$n$ is an integer, start to be observable in the $IV$
characteristic \cite{us03}.

These results correspond to the fully phase-coherent case ($\eta=0$) and,
as can be observed, exhibit a highly non-linear behavior where a linear
regime is strictly never reached. The presence of a finite $\eta$ in the
actual system limits the observation of this behavior, giving
rise to a linear regime in the limit $V \rightarrow 0$. This is
illustrated in the right panel of Fig. \ref{ivteorica} where the $IV$
characteristic for $\Gamma^* = 2 \Delta$ is represented for different
values of $\eta$. As can be observed, the current at low bias is
extremely sensitive to the actual value of this parameter. Notice also
that the oscillations with period $\Delta/n$ are suppressed for $eV \le
\eta$.

These results can be understood more deeply by analyzing the dynamics of
the Andreev states in the presence of an applied bias taking into
account the finite relaxation rate. Let us first consider the symmetric
case. In the limit $eV < \eta$ we can
obtain the non-equilibrium occupation of the Andreev states, $\rho^{\pm}$,
using the Boltzman equation in the relaxation time approximation, i.e.

\begin{eqnarray}
\frac{\partial \rho^{\pm}}{\partial t} =
-\frac{\rho^{\pm} - \rho^{(0)\pm}}{\tau}
\end{eqnarray}
where $\rho^{(0)\pm} = n_F\left[\pm\omega_S(\phi)\right]$ and $\tau =
\hbar/2\eta$.
In the stationary state these quantities can be expanded as
$\rho = \sum_n \rho_n \exp{in\omega_0t}$, where $\omega_0 = eV/\hbar$
and we obtain

\begin{eqnarray}
\rho^{\pm}_n = \frac{\rho^{(0)\pm}_n}{1 + i n \omega_0 \tau}
\end{eqnarray}

At zero temperature the Fourier components of the equilibrium
distribution are simply given by $\rho^{(0)\pm}_{2k+1} = (-1)^k/(\pi
(2k+1))$. To
calculate the current we first notice that within this approximation

\begin{eqnarray}
I = \frac{e}{\hbar} \frac{\partial \omega_S}{\partial \phi}
\left[\rho_+ - \rho_- \right]
\end{eqnarray}

Averaging this expression to obtain the dc current,
assuming that $\omega_S \simeq \widetilde{\Delta} \cos{\phi/2}$, yields

\begin{eqnarray}
I_0 = \frac{2e^2\widetilde{\Delta}}{h\eta} \frac{V}{1 +
(eV/2\eta)^2}
\label{analytical-dc}
\end{eqnarray}

This expression indicates that the conductance reaches a maximum
at $V=0$ where it takes the value $G(0) = 2e^2 \widetilde{\Delta}/(h \eta)$
and decreases with $V$ defining a zero-bias conductance peak
with a typical width fixed by $\sim \eta$.
In panel (b) of Fig. \ref{ivteorica} we compare the results of
this approximation
with the full numerical ones. As can be observed Eq. (\ref{analytical-dc})
describes correctly the behavior for $eV < \eta$.

This description can be straightforwardly generalized to a non-symmetric
situation. In this case the effect of the transition between the lower
and the upper state can be analyzed as a Landau-Zener process with
probability $p = \exp{\left(-\pi r \Delta/eV\right)}$, where $r = 1 - \tau$
\cite{averin}. The current in Eq. (\ref{analytical-dc}) will be then
suppressed by a factor $p$.

On the other hand for finite temperature $T$,
we obtain $I_0(T) = I_0(T=0)F(T)$, where $I_0(T=0)$ is given by Eq.
(\ref{analytical-dc}) and

\begin{eqnarray}
F(T) = \frac{2\pi}{\beta \widetilde{\Delta}} \sum_{m=0}^{\infty}
\left(1 - \frac{x_m}{\sqrt{1 + x_m^2}} \right)
\end{eqnarray}
where $\beta = 1/k_B T$ and $x_m = (2m+1)\pi/(\beta\widetilde{\Delta})$.
The absence of any dependence on $\eta$ or $V$ in $F(T)$
indicates that the main effect of temperature is to introduce a
global scale factor reducing the magnitude of the conductance.

\begin{figure}
\includegraphics[width =\columnwidth]{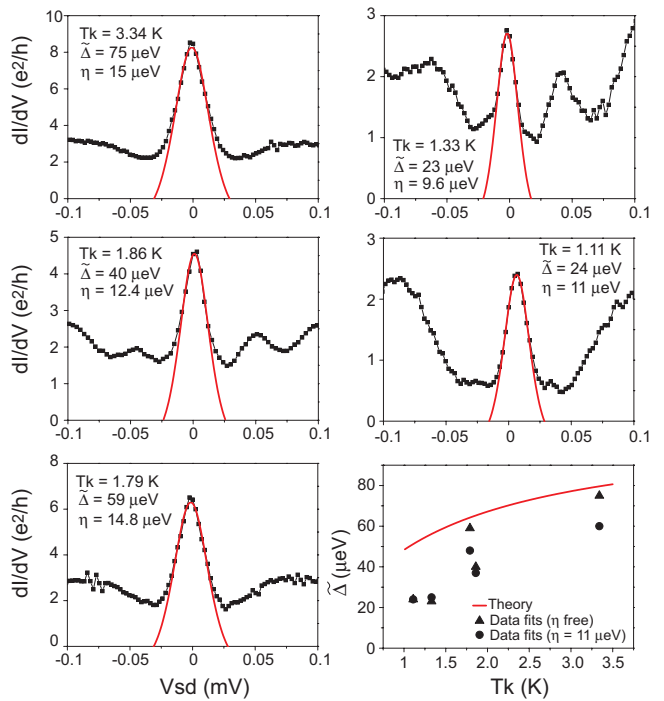}
\caption{Differential conductance in the middle of 5 different
Kondo ridges with decreasing $T_K$. The red lines are fits to the
data using Eq.\ref{analytical-dc}. The lower right pannel shows
the fitted values of $\widetilde{\Delta}$ vs $T_K$ compared to the
theoretical prediction $\widetilde{\Delta}/\Delta \simeq 1/(1 +
\Delta/T_K)$.} \label{peaks}
\end{figure}

This simple theory can be used to analyze the low bias conductance peak
in the experimental results. As shown in Fig. \ref{peaks}, this peak can
be well fitted by the expression (\ref{analytical-dc}) using $\eta$ and
$\widetilde{\Delta}$ as free parameters. The values of $\eta$ thus
obtained are rather constant (between $\sim 10$ and $\sim 15 \mu eV$) while
$\widetilde{\Delta}$ strongly depends on $T_K$. This behavior is
consistent with our theoretical model in which the inelastic relaxation
rate is a fixed parameter set by some microscopic mechanism not
explicitly included in the model. Notice that $\eta$ is considerable
smaller than both $\Delta$ and $\delta \epsilon$. 
On the other hand the
theory predicts an increase of $\widetilde{\Delta}$ as a function of $T_K$
roughly given by $\widetilde{\Delta}/\Delta \simeq 1/(1 + \Delta/T_K)$.
In the lower right pannel of Fig. \ref{peaks} this prediction is
compared with the fitted experimental values. Notice that, in spite of
the rather large dispersion in the data, both the order of magnitude and the
general trend are in agreement with the theoretical predictions.

In conclusion we have presented a combined theoretical and
experimental analysis of electron transport in carbon nanotubes
coupled to superconducting leads. The differential conductance in
the Kondo regime exhibits a zero bias peak which can be related
to the dynamics of the subgap Andreev states under an applied
bias voltage. A main ingredient that has to be included in the
model in order to account for the experimental results is a finite
relaxation rate $\eta$ which gives rise to a linear regime in the
low bias limit $eV < \eta$. The different mechanisms that can
contribute to this finite rate can be either intrinsic (like
electron-phonon coupling within the nanotube) or extrinsic (for
instance due to phase diffusion like in a RSJ model
\cite{kang03}). Further experiments could be of interest in order
to elucidate this issue.

\acknowledgments
The authors would like to thank J.C. Cuevas for useful discussions.
Work was supported by EU through DIENOW Research Training Network.

\noindent
$^*$ Present address: Cavendish Laboratory, Cambridge, UK.


\begin{thebibliography}{}
\bibitem{comment:slave-boson} An equivalent description in terms of
renormalized parameters is obtained
within the slave-boson mean field approach. See for instance last reference
in \cite{avishai}.
\end{thebibliography}
\end{document}